\documentclass[twocolumn,preprintnumbers,amsmath,amssymb]{revtex4}

\usepackage{graphicx}
\usepackage{dcolumn}
\usepackage{bm}
\usepackage{colordvi}
\usepackage{epsfig}
\usepackage{graphicx}
\usepackage{color}

\newcommand{\kap}{\boldsymbol{\kappa}}
\newcommand{\rb}{{\bf r}}
\newcommand{\bp}{\boldsymbol{\partial}}
\newcommand{\qb}{{\bf q}}
\newcommand{\kb}{{\bf k}}

\newcommand{\sig}{\boldsymbol{\sigma}}

\newcommand{\F}{\boldsymbol{F}}
\newcommand{\B}{\boldsymbol{B}}
\newcommand{\C}{\boldsymbol{C}}
\newcommand{\R}{\boldsymbol{R}}

\begin{document}

\title{A First-Principles Constitutive Equation for Suspension Rheology: \\
Supplementary Material}

\author{J.~M.~Brader$^1$, M.~E.~Cates$^2$ and M.~Fuchs$^1$}
\affiliation{
$^1$Fachbereich Physik, Universit\"at Konstanz, D-78457 Konstanz, Germany
\\
$^2$SUPA, School of Physics, The University of Edinburgh,
Mayfield Road, Edinburgh EH9 3JZ, UK
}

\maketitle

In this supplementary material we first provide additional details of 
our integration through transients approach to the constitutive 
equation and distorted microstructure, clarifying the approximations 
required to obtain Eqs.(11) and (12). 
We then present an exact result for the friction kernel 
and details of the approximations leading to Eqs.(13)-(15).
Finally, we provide additional comments regarding two continuum mechanics 
principles from the standpoint of the present work.

\section{Integration Through Transients}
Central to our integration through transients approach is Eq.(10) which 
enables averages of functions of the particle coordinates to be calculated 
[7]. 
Applying (10) to a general function $f$ of the phase space 
coordinates yields the time-dependent average
\begin{eqnarray}
\langle f \rangle(t) 
= \langle f \rangle + \int_{-\infty}^{t}\!\!dt' 
\langle{\rm Tr}\{\kap(t')\hat{\boldsymbol{\sigma}}\}
e_-^{\int_{t'}^t ds\,\Omega^{\dagger}(s)}
f \rangle.
\notag
\end{eqnarray}
The time-ordered exponential function $e_{-}$ generalizes the  
exponential function for (differential) operators \cite{vankampen}.
Choosing $f=\hat\sigma_{\alpha\beta}/V$, where $V$ is the system volume, yields 
directly an exact generalized Green-Kubo relation for the stress tensor
\begin{equation}
{\boldsymbol \sigma}(t) = \frac{1}{V} \int_{-\infty}^{t}\!\!\!\!\! dt'\,
\langle  
{\rm Tr}\{\kap(t')\hat{\boldsymbol{\sigma}}\} 
e_-^{\int_{t'}^t ds\,\Omega^{\dagger}(s)} \hat{\boldsymbol \sigma}\rangle.
\notag
\label{exact_stress}
\end{equation}
This equation is nonlinear in $\kap(t)$ due to the appearance of  
$\Omega^{\dagger}(t)$ in the exponent. 
The stress autocorrelation function $\langle\,\cdot\,\rangle$ appearing 
in the above expression is the overlap between a stress fluctuation at 
time $t'$ and a second fluctuation at later time $t$ which has evolved under a 
combination of shear-driven and internal motion. 
This correlation function is approximated by calculating overlaps of the stress fluctuations 
with density fluctuations describing slow structural relaxation of the dense 
system. 
The lowest order non-zero projection of the dynamics is onto pairs of 
density fluctuations and is achieved using the pair projector
\begin{eqnarray}
P_{2}=\sum_{{\bf k}>{\bf p}}\rho^{}_{{\bf k}}\rho^{}_{{\bf p}}\rangle 
\frac{1}{N^2 S_{k}S_{p}}
\langle\,\rho^{*}_{{\bf k}}\rho^{*}_{{\bf p}}.
\notag
\end{eqnarray}
Insertion of this projector either side of the time-ordered exponential 
function followed by factorization of a fourth moment in density fluctuations 
into the product of second moments (viz. the correlators)
leads to an approximate expression for the stress tensor.  
The time evolution of stress fluctuations is thus represented 
by the transient density correlator $\Phi_{\bf k}(t,t')$. 
Within our approach the static structure factor $S_k$ serves to represent 
the direct potential interactions. 
For glassy states ageing leads to a residual time dependence of $S_k$ which 
cannot be resolved within our approach. 
We therefore employ an extrapolation of the equilibrium $S_k$ when working 
above the glass transition.
\par
In order to calculate the distorted structure factor we use our general 
result (10) to calculate the average of
$\Delta\rho^*_{\bf k}\rho^{}_{\bf k}=\rho^*_{\bf k}\rho^{}_{\bf k} 
- \langle\rho^*_{\bf k}\rho^{}_{\bf k}\rangle$
\begin{eqnarray}
S_{\bf k}(t;\kap)
\!=\! \langle \rho^*_{\bf k}\rho^{}_{\bf k} \rangle \!+\! \int_{-\infty}^{t}\!\!dt' 
\langle{\rm Tr}\{\kap(t')\hat{\boldsymbol{\sigma}}\}
e_-^{\int_{t'}^t ds\,\Omega^{\dagger}(s)}
\Delta\rho^*_{\bf k}\rho^{}_{\bf k} \rangle.
\notag
\end{eqnarray}
As for the stress we can apply the projection operator $P_2$ to approximate 
the average in the integrand, leading to the expression (11) for the 
distorted structure factor. 
We note that in calculating $S_{\bf k}(t;\kap)$ we require Baxter's 
result 
$
\langle \rho^{*}_{\bf k}\rho^{}_{\bf k}\rho^{}_{\bf 0} \rangle = 
NS_0 \left( S_k + n\frac{\partial S_k}{\partial n} \right)
$
to treat the three point function which arises as a result of the projection 
operator steps \cite{baxter}.

\section{Equation of Motion}
In order to close our constitutive theory we require an equation of motion 
for the transient density correlator. 
The advected wavevector appearing in the definition of $\Phi_{\bf k}(t,t')$ 
may be reformulated as   
\begin{eqnarray}
\Phi_{\bf k}(t,t')&=&
\frac{1}{N S_{k}}
\langle\,\rho^*_{\bf k}
e_{-}^{\int_{t'}^t ds\,\Omega^{\dagger}(s)}
\rho_{\bar{\kb}(t,t')} \rangle \notag\\
&=&
\frac{1}{N S_{k}}
\langle\,\rho^*_{\bf k}
e_{-}^{\int_{t'}^t ds\,\Omega^{\dagger}(s)}
e_{-}^{-\int_{t'}^t ds\,\delta\Omega^{\dagger}(s)}
\rho_{{\kb}} \rangle
\notag
\end{eqnarray}
where we introduce 
$\delta\Omega^{\dagger}(t) = \sum_i\rb_i\cdot\kap^T(t)\cdot\bp_i$.
Exact operator identities generalizing [7] lead directly to the 
first equation of motion
\begin{eqnarray}
\frac{\partial}{\partial t}\Phi_{\bf q}(t,t_0) &+& 
\Gamma_{\bf q}(t,t_0)\Phi_{\bf q}(t,t_0) \notag\\
&&\hspace*{-0.3cm}+ \int_{t_0}^{t}dt' \,M_{\bf q}(t,t',t_0)\Phi_{\bf q}(t',t_0) 
=\Delta_{\bf q}(t,t_0).
\notag
\end{eqnarray}
The initial decay rate is given explicitly by 
$\Gamma_{\bf q}(t,t_0)=\bar{q}^2(t,t_0)/S_{\bar{q}(t,t_0)}$.
Exact formal expressions are obtained for both the generalized diffusion kernel 
$M_{\bf q}(t,t',t_0)$ and correction function $\Delta_{\bf q}(t,t_0)$. 
Regarding the first equation of motion as a Volterra integral equation
of the second kind enables the final exact form for the 
equation of motion to be formulated
\begin{eqnarray}
\hspace*{0.cm}
\frac{\partial}{\partial t}\Phi_{\bf q}(t,t_0)
&+& \Gamma_{\bf q}(t,t_0)\bigg(
\Phi_{\qb}(t,t_0)
\notag\\
&&\hspace*{-1cm}+
\int_{t_0}^t dt' \,m_{\qb}(t,t',t_0) \frac{\partial}{\partial t'} \Phi_{\qb}(t',t_0)
\bigg) = \tilde\Delta_{\bf q}(t,t_0).
\notag
\end{eqnarray} 
\par
The generalized friction kernel and modified correction function are given formally by 
\begin{eqnarray}
m_{\bf q}(t,t',t_0)&=& 
\frac{
\langle
\rho^*_{\bf q} \Omega_a^{\dagger}(t',t_0) U_i(t,t',t_0)
\Omega^{\dagger}_r(t,t_0)\rho_{\bf q}
\rangle}
{
NS_{\bar{q}(t',t_0)} \Gamma_{\bf q}(t',t_0) \Gamma_{\bf q}(t,t_0)
},\notag\\
\tilde\Delta_{\bf q}(t,t')&=&
\frac{\langle \rho^{*}_{\bf q} U_i(t,t',0)\,\Omega^{\dagger}_r(t)\rho^{}_{\bf q} \rangle}{NS_q},
\notag
\end{eqnarray}
where we have introduced the operators
\begin{eqnarray}
\Omega_a^{\dagger}(t',t_0)&=&
e_-^{\int_{t_0}^{t'} ds \,\overline{\delta\Omega}^{\dagger}(s)}
\Omega_e^{\dagger}\,
e_+^{-\int_{t_0}^{t'} ds \,\delta\Omega^{\dagger}(s)},
\notag\\
\Omega^{\dagger}_r(t,t_0)&=&
e_-^{\int_{t_0}^{t} ds \,\delta\Omega^{\dagger}(s)}
Q(t,t_0)\Omega^{\dagger}_e\,
e_-^{-\int_{t_0}^{t} ds \,\delta\Omega^{\dagger}(s)},\notag\\
\Omega_i^{\dagger}(t,t_0) &=& \Omega^{\dagger}_r(t,t_0)
\left( 1 
- \frac{\rho^{}_{\bf q}\rangle\langle\,\rho^{*}_{\bf q}\Omega_a^{\dagger}(t,t_0)} 
{\langle \,\rho^{*}_{\bf q}\,\Omega_{a}^{\dagger}(t,t_0)\,\rho^{}_{\bf q} \rangle}\right),
\notag
\end{eqnarray}
which depend upon $\overline{\delta\Omega}^{\dagger}(t)=
\sum_i\rb_i\cdot\kap^T(t)\cdot(\bp_i + {\bf F}_i)$ and
$Q(t,t_0) \!=\! 1 \!-\! \sum_{\bf q}
\rho^{}_{\bar{\bf q}(t,t_0)}\,\rangle
\frac{1}{NS_{\bar{q}(t,t_0)}}
\langle\,\rho^{*}_{\bar{\bf q}(t,t_0)}$.
The irreducible part of the dynamics is contained in 
\begin{eqnarray}
U_i(t,t',t_0)=
e_-^{\int_{t'}^{t} ds \,\Omega_i^{\dagger}(s,t_0)}.
\notag
\end{eqnarray}
\par
All results are at this stage formally exact.
Approximation of the friction kernel 
proceeds via two steps and is based on the assumption that $U_{i}(t,t',t_0)$ contains 
slow dynamics only because of coupling to higher density modes describing structural relaxation. 
Firstly, we project the average in the numerator
onto density pairs using the time dependent projection operator 
\begin{eqnarray}
P_{2}(t,t_0)=\sum_{{\bf k}>{\bf p}}\frac{\rho^{}_{\bf \bar{k}(t,t_0)}\rho^{}_{\bf \bar{p}(t,t_0)}\rangle 
\langle\,\rho^{*}_{{\bf \bar{k}}(t,t_0)}\rho^{*}_{{\bf \bar{p}(t,t_0)}}}
{N^2 S_{\bar{k}(t,t_0)}S^{}_{\bar{p}(t,t_0)}}.
\notag
\end{eqnarray} 
This reduces the problem to the calculation of a four point correlation function. 
The second step is to approximate this correlator, 
in the spirit of quiescent mode coupling theory, by a product of pair correlators. 
The modified correction function makes a negligable contribution for small 
accumulated strains which suggests the approximation 
$\tilde\Delta(t,t_0)\!=\!0$ \cite{future}.
We thus arrive at Eqs.(14) and (15) given in the main text.\\

\section{Continuum Mechanics Principles}

The principle of material objectivity states that the relationship between 
the stress and strain tensors should be independent of the rotational state 
of either the sample or the observer 
[11,16,17]. 
This is satisfied by the Smoluchowski equation which neglects inertial effects. 
That this invariance is preserved in our approximate equations can be explicitly 
confirmed by considering the imposition of a time dependent rotation onto an 
arbitrary flow. The shear gradient, deformation gradient, left and right 
Cauchy-Green tensors in the rotating frame are thus given by
\begin{eqnarray}
\hat{\kap}(t) &=& \R(t)\kap(t)\R^T(t)+\dot{\R}(t)\R^T(t)\notag\\
\hat{\F}(t,t')&=& \R(t)\F(t,t')\R^T(t')\notag\\
\hat{\B}(t,t')&=& \R(t)\B(t,t')\R^T(t)\notag\\
\hat{\C}^{-1}(t,t')&=& \R(t')\C^{-1}(t,t')\R^T(t'),\notag
\end{eqnarray}
where $\R(t)$ is a time-dependent rotation matrix. 
It is a straightforward but laborious exercise to substitute the transformed 
tensors into expressions (8),(9),(11-14) in order to obtain the required 
invariance result for the stress tensor
$\hat{\sig}(t)=\R(t)\,\sig(t)\R^{T}(t).$
\par
An invariance requirement often implicitly employed in continuum modeling is Oldroyd's
{\em principle of local action} [11] which states that only 
neighbouring particles are involved in determining the stress at any given point.
The ${\bf k}$ integrals in (11) and (12) are insensitive to the low-$k$ behavior 
of the integrands. 
In this sense local action is substantiated by the present microscopic theory.

\end{document}